%% file: main.tex
\documentclass[journal]{IEEEtran}
\IEEEoverridecommandlockouts

\usepackage[english]{babel}

\usepackage{graphicx}
\usepackage[colorlinks=true, allcolors=blue]{hyperref}
\usepackage{epsfig}
\usepackage{times} 
\usepackage[cmex10]{amsmath}
\usepackage{bm}
\usepackage{algorithmic}
\usepackage{algorithm}
\usepackage{caption} 
\usepackage{subcaption} 
\usepackage[dvipsnames]{xcolor}
\usepackage{multirow}
\usepackage{cite}
\usepackage{svg}
\usepackage{balance}

\newfloat{model}{thp}{lop}
\floatname{model}{Model}

\renewcommand{\baselinestretch}{0.98}

\title{Security Constrained Optimal Power Shutoff}

\author{Noah Rhodes,
Carleton Coffrin,
~Line Roald
\thanks{Noah Rhodes and Line Roald are with the Department of Electrical and Computer Engineering, University of Wisconsin, Madison, WI, USA. Their work is supported by the U.S. National Science Foundation ASCENT program under award 2132904 and by the Department of Energy, Office of Science, Office of Advanced Scientific Computing Research, Applied Mathematics program under Contract Number DE-AC02-06CH11347.}%
\thanks{Noah Rhodes and Carleton Coffrin are with the Los Alamos National Laboratory, Los Alamos NM, USA.}%
}

\begin{document}
\maketitle

\begin{abstract}

Electric grid faults are increasingly the source of ignition for major wildfires.
To reduce the likelihood of such ignitions in high risk situations, utilities use pre-emptive deenergization of power lines, commonly referred to as Public Safety Power Shut-offs (PSPS).
Besides raising challenging trade-offs between power outages and wildfire safety, PSPS removes redundancy from the network just at a time when component faults are likely to happen. This may leave the network particularly vulnerable to %
unexpected line faults that may occur while the PSPS is in place. Previous works have not explicitly considered the impacts of such outages.
To address this gap, we propose the \emph{Security-Constrained Optimal Power Shutoff} (SC-OPS) problem which uses post-contingency security constraints to model the impact of unexpected line faults when planning a PSPS.
This SC-OPS model enables, for the first time, the exploration of a wide range of trade-offs between both wildfire risk and pre- and post-contingency load shedding while designing PSPS plans, providing useful insights for utilities and policy makers considering different approaches to PSPS.
We demonstrate the efficacy of our model using the EPRI 39-bus test system as a case study. %
The results highlight the potential risks of not considering security constraints when planning PSPS and 
show that incorporating security constraints into the PSPS design process improves the resilience of current PSPS plans.

\end{abstract}

\section{Introduction}
In recent years, the number and severity of wildfires ignited by electric power lines have increased. Several of the %
most damaging fires in California from 2018-2021 were ignited by power lines, including
the Camp Fire in 2018 \cite{porter_2018}, the Kincade Fire in 2019 \cite{porter_2019}, and the Dixie Fire in 2021 \cite{tyler_2021}.  
The number of fires ignited by power lines in California and their destruction has increased in recent years \cite{miller2017electrically, keeley2018historical}. From 1960-2009, power lines represented 1.1\% of all wildfires \cite{mitchell2013power}.  From 2015-2020, the power grid accounted for 10\% of wildfire ignitions, and  70\% of all damages from wildfires (\$17.5 billion) \cite{noauthor_stats_nodate}. 

A major challenge for power system operators is that many of the same weather conditions that increase wildfire spread also increase rates of component failures.  Weather impacts like high temperatures, low humidity, and high wind speeds can cause a small ignition to rapidly spread into a large and uncontrollable wildfire.  The same weather increases power line fault rates. For example, high temperatures increase line sag through thermal expansion, allowing power lines to potentially touch nearby vegetation, and wind speed can cause conductor slap when the lines are pushed into each other in high wind. Overall utility outage rates increase 10,000 fold with 95 km/hr wind speeds \cite{mitchell2013power}.  

When planning a Public Safety Power Shutoff (PSPS), the highest risk lines, those which have an elevated probability of a fault and an increased probability of an ignition spreading into a large damaging wildfire, are de-energized.  However, this does not mean that the remaining energized lines will not fault.  Prior to the 2019 Kincade Fire in Sonoma County, California, a PSPS shut off multiple distribution circuits throughout the state, including in Sonoma County \cite{pgeOct2019report}. However, nearby transmission lines were not de-energized.  The planned PSPS began at 12:00pm and disconnected customers on the distribution system, but at 9:20pm a fault on a nearby 230 kV energized transmission line \cite{cpuc_investigation_08_23_2019} would become the ignition source for the largest wildfire in California that year \cite{porter_2019}.  

During a PSPS event, 
the remaining energized lines can still experience a fault or line failure, which can result in a dangerous wildfire or increased customer outages.
For this reason, it is important to consider the impacts of contingency events on a PSPS plan by investigating the tradeoff between load shed, wildfire risk, and system reliability.

\subsection{Related Work}
The increase in wildfire ignitions caused by power infrastructure resulted in a shift of research focus from the impacts of wildfires on the power grid, as seen in \cite{7995099, CHOOBINEH201520, mohagheghi2015optimal, moutis2022pmu, nazemi2022powering, dian2019integrating, sathaye2013rising}, to also considering the impact of the power grid on wildfire ignitions. Refs. \cite{8768218, 8767948, arab2021three, muhs2020wildfire,vazquez2022wildfire, nazaripouya2020power} review approaches to reduce the risk of ignitions, many of which are already being implemented or planned by utilities.
There are important infrastructural improvements that are necessary in the long-term to reduce this risk, such as placing more power lines underground and improved vegetation clearing \cite{noauthor_pge_nodate}, but these investments take years or decades to implement.  

Short-term operational changes can be made to reduce the wildfire risk, the most effective and most disruptive of which is de-energization of power lines \cite{noauthor_utility_nodate}.  
This has been implemented by California utilities including Pacific Gas \& Electric (PG\&E) under the name \emph{Public Safety Power Shutoff} (PSPS) where equipment in regions with high wildfire risk is de-energized \cite{noauthor_psps_nodate}. In PG\&E's PSPS from October 9-12th 2019, 100 instances of damage were found on de-energized lines, pointing to evidence of wildfire prevention, however, 3 million people lost power during the event \cite{pge_factsheet}.

Previous work has investigated approaches to optimize a de-energization plan that minimizes wildfire ignition risk while maximizing the load demand that can be met \cite{lesage2022optimally, rhodes2020balancing}.  Data driven and machine learning methods attempt to accelerate the solution time of these optimization problems, which can be very difficult to solve to optimality \cite{hong2022data, umunnakwe2022data, bayani2022quantifying}. 
Other research uses the Optimal Power Shutoff problem as a basis to plan infrastructure upgrades that minimize the impact of PSPS or reduce the need for PSPS in the future. Resilience through battery installations or under-grounding of lines is studied in \cite{kody2022optimizing, taylor2022framework, bertoletti2022transmission, bayani2023resilient}, resilience through microgrid operations in \cite{yang2022resilient, hanna2021optimal}, and load shifting through energy storage to reduce power outages in \cite{astudillo2022managing}. 
Other OPS research considers extensions necessary for distribution grids \cite{gorka2022efficient}, concurrent planning of PSPS and grid restoration \cite{rhodes2022cooptimization}, considerations of fairness for load outages \cite{kody2022sharing}, improved forecasting of wildfire ignition risk \cite{yao2022predicting, bayani2022quantifying}, and dynamic line rating to reduce current rather than fully de-energize lines \cite{tandon2021motivating}.

However, an important aspect of planning a PSPS that is not very well addressed is the resiliency of the resulting network after a PSPS is conducted.  
In \cite{rhodes2022cooptimization} authors consider a `vulnerability threshold' to penalize de-energization of low-risk lines and maintain some redundancy in the transmission network. Refs. \cite{CHOOBINEH201520, hong2022data, rhodes2020balancing, umunnakwe2022data, kody2022optimizing, astudillo2022managing, rhodes2022cooptimization, nazaripouya2020power, yang2021criterion, tandon2021motivating, mohagheghi2015optimal} all note the need for security constraints to ensure safe operations in future work, however, none of these works include such security constraints on grid operations.

The goal of security constraints is to ensure the safe operation of the network if a component fails \cite{alsac1974optimal}.  To be secure against N-1 contingencies in normal operations, power flow must be feasible if any single power line fails \cite{capitanescu2011state}.  The resulting optimization problem, typically referred to as Security Constrained Optimal Power Flow, is often modeled using sensitivity factors to model the transfer of power from a line that fails to the rest of the network \cite{stott2009dc}.  However, in modeling a PSPS, we do not know the topology of the network before solving the problem, and thus do not know the sensitivity factors. In addition, since PSPS leads to reduced grid redundancy, we must consider that contingencies can cause the splitting of the grid into several islands. Research on transmission switching has accounted for the former, but not the latter \cite{schnyder1988integrated}.

\subsection{Contributions}
The contributions of this work are as follows:
\begin{enumerate}
    \item We propose a model for security-constrained optimal power shut-off (SC-OPS). This model extends prior work by including several contingency scenarios in the optimization model. We also design the objective function to minimize wildfire risk, and introduce explicit bounds on the amount of pre- and post-contingency load shed we allow.
    \item Through a case study on the IEEE RTS system, we perform an analysis of the trade-offs associated with introducing security constraints, including impacts on the size of the resulting power outage for a given level of acceptable wildfire risk, how additional generation flexibility can reduce power outages and fire risk, and the trade-off between pre- and post-contingency load shed.
    \item Release of the models proposed in this work in the open-source software package \emph{PowerModelsWildfire.jl}.
\end{enumerate}

\subsection{Paper Organization}
The paper is organized as follows:
Section \ref{sec:vulnerability} discusses how the power grid becomes more vulnerable during high wildfire risk events. 
Section \ref{sec:modeling} introduces the three mathematical models used in this work.
Section \ref{sec:casestudy} describes the case study.
Section \ref{sec:results} presents the results from the case study analysis and
Section \ref{sec:discussion} discusses some key impacts of the results. 
Finally, section \ref{sec:conclusion} concludes the work.

\section{Operational Security During Public Safety Power Shutoff} \label{sec:vulnerability}
The goal of implementing a PSPS is to reduce the likelihood of a component failure in the power grid causing an ignition that sparks a large and damaging wildfire.  This is an example of a \emph{high-impact, low-probability event} where the probability of any line causing an ignition is low, but the impact of a major wildfire is catastrophic.  Managing the risk of wildfire ignitions under these conditions is challenging.  

Implementing a power shutoff plan results in a certainty of some components being de-energized (and some customers losing power) instead of the uncertainty that a component in a high risk region faults and ignites a wildfire.  The more components that are de-energized results in more certainty about the impacts, e.g. de-energizing the entire grid will result in all customers losing power, but ensures that no component failure is able to ignite a wildfire.  
Planning a PSPS requires the system operator to balance these certain and uncertain impacts. The goal is to de-energize the highest risk components that are most likely to have failures and to ignite the most destructive wildfires, without causing widespread customer power outages.

The best way to reduce wildfire risk without shedding load is to de-energize all the redundant components that pose a risk of causing a fault. 
As an example, a solution to the Optimal Power Shutoff Problem that optimizes a maximum decrease in wildfire risk with a minimal decrease in load delivery often results in a fully radial network \cite{rhodes2020balancing}. 
This happens because redundant components typically contribute to increased wildfire ignition risk without improving load delivery.
Unfortunately, removing redundancy means that any additional fault can result in a major load outage. 
This is particularly important to consider because PSPS happen during a 
time where we already expect higher rates of faults due to weather conditions like high wind speed.  This is not a contrived scenario, the 2019 Kincade fire is an example of a fault and wildfire ignition occurring after a PSPS was implemented \cite{pgeOct2019report}.

The challenge in implementing a PSPS is to balance power outages with wildfire ignition risk. While some outages due to PSPS are easy to assess (i.e., they will happen with certainty) it is also necessary to consider the load shed implications of additional outages due to faults. Incorporating the consideration of N-1 security constraints is a step towards a more comprehensive assessment of load shed. N-1 security is an accepted reliability criterion in normal operations, and can provide a similar benefit to PSPS as well.  However, the chief goal of N-1 security constraints are to prevent load shed in normal operation, while the PSPS permits load shed to reduce wildfire risk. 
This leads to important distinctions in the modeling, which are adressed in this paper as described below.
In this work, we show that incorporating security constraints in PSPS planing  is more resilient and reduces the number of customers that may experience a power outage.

\section{Modeling the Security-Constrained Optimal Power Shutoff Problem} \label{sec:modeling}
This section describes the mathematical modeling of the SC-OPS problem.  We consider a network with buses $ i \in \mathcal{B}$, generators $ g \in \mathcal{G}$, lines $ij \in \mathcal{L}$, and load $d \in \mathcal{D}$. The contingency set contains line outage scenarios  $c \in \mathcal{C}$, where $\mathcal{L}_c$ is the set of lines in the contingency scenario $c$. The pre-contingency scenario is $\mathcal{C}_0$ where the only element is $c=0$.  Parameters are bold values while variables are non-bold.  Binary variables representing the energization state of a component are $z$ and indexed according to the component.  Power flow is modeled using linearized DC power flow equations, with line switching constraints allowing lines to be de-energized.

\noindent \emph{Optimization Objective:}
The objective function of the SC-OPS problem \eqref{eq:ops_objective} seeks to minimize the wildfire risk in the power grid, where energized lines contribute an amount of risk $\boldsymbol{R}^L_{ij}$ associated with each line $ij \in \mathcal{L}$ and de-energized lines contribute no risk.  The state of the line $ij$ in contingency scenario $c$ is represented by the binary variable $z_{ijc}$  where $z_{ijc}=1$ indicates that the line is energized, and $z_{ijc}=0$  indicates that the line is de-energized.  The total wildfire risk of the network in the pre-contingency scenario is then the sum of the wildfire risk of each energized line  when $c=0$. 
\begin{equation}
        min \sum_{(i,j)\in \mathcal{L}} z^L_{i j 0} \boldsymbol{R}^L_{i j}
        \label{eq:ops_objective}
\end{equation}
The minimum risk solution is to de-energize the entire grid, therefore we add a constraint \eqref{eq:ops_load_threshold} to limit the amount of load shed in the pre-contingency scenario.
Variable $x_{d0} \in [0,1]$ represents the portion of load demand $\boldsymbol{D}_{d}$ that is served.  The summation of $x_{d0} \boldsymbol{D}_d$ over all loads $d \in \boldsymbol{D}$  is the total load served in the pre-contingency scenario.  This amount must be greater than or equal to $\boldsymbol{\alpha} \boldsymbol{D}^{Tot}$ where $\boldsymbol{D}^{Tot}$ is the total load demand of the system $\sum_{d\in\mathcal{D}}\boldsymbol{D}_d$ and $\boldsymbol{\alpha}$ is an input parameter between 0 and 1 that dictates the required amount of load served.
For example, we may set $\boldsymbol{\alpha} = 0.9$ to require
the load served to be at minimum 90\% of the total system load demand and find the operating point that minimizes wildfire risk while satisfying this constraint. 
\begin{equation}
        \sum_{d\in\mathcal{D}}x_{d0} \boldsymbol{D}_d \ge \boldsymbol{\alpha}{\boldsymbol{D}^{Tot}} \label{eq:ops_load_threshold}
\end{equation}
This formulation is an easy tool for policy-makers to use because it can directly answer the question: \emph{If we shed 5\% of system load, how much wildfire risk will this reduce?} It also enables a direct comparison of wildfire risk reduction as the input parameters are adjusted or the power system is modified. 

\noindent \emph{Pre-contingency Operational Constraints:} ($\forall c \in \mathcal{C}_0$)
The following constraints all apply to the pre-contingency scenario $c=0$.
Equations in \eqref{eq:ops_gen_active}-\eqref{eq:ops_demand_active} prescribe how de-energized components can be connected.  For any generator $g$, load $d$, or line $ij$, if the connected bus $i$ is de-energized, $z_{i0}=0$, then the component must also be de-energized.  For example, for each load connected to bus $i$, $d \in \mathcal{B}^\mathcal{D}_i$, if $z_{i0}=0$ then $x_{d0}=0$.  If the bus is energized, the load can be served, and $x_{d0}$ can take any value in its range $[0,1]$.  The same is true for generators and lines, except their state is a binary value.
\begin{subequations}
\begin{align}
        &  x_{d0} \le z_{i0}  &&   \forall d \in \mathcal{B}^\mathcal{D}_i, \; \forall i \in \mathcal{B}  \label{eq:ops_gen_active} \\
        &  z_{ig} \le z_{i0}  &&  \forall g \in \mathcal{B}^\mathcal{G}_i, \; \forall i \in \mathcal{B}   \label{eq:ops_line_active}\\
        &  z_{i j 0} \le z_{i0} &&  \forall (i, j) \in \mathcal{B}_i^{\mathcal{L}}, \; \forall i \in \mathcal{B}   \label{eq:ops_demand_active}
\end{align}
\label{eq:ops_relationships}
\end{subequations}
\vspace{-1.2em} %

\noindent \emph{Post-contingency Operational Constraints:} ($\forall c \in \mathcal{C}$)
The following constraints are enforced for each line outage contingency $c \in \mathcal{C}$.  First are the energization states for each component.  
Constraint \eqref{eq:scops_line_contingency} requires any line in the contingency $c$ to be de-energized $z_{i j c}=0$.  
Each other line in the network must have the same energization state as in the pre-contingency case $z_{i j c} = z_{i j 0}$, and is enforced by \eqref{eq:scops_line_active}. 
Constraint \eqref{eq:scops_bus_active} requires that the post-contingency energization state of each bus is the same as the pre-contingency energization state.  
In Eq. \eqref{eq:scops_gen_active}, a generator is permitted to become de-energized post-contingency, modeling a generator tripping offline if a change in power flow exceeds the limitations of the generator response.  
Finally, \eqref{eq:scops_no_pickup} models the load state. The load served post-contingency $x_{dc}$ cannot increase because it is not possible to re-energize loads that were shut off to prevent fires (damaged components could cause ignitions when energized), but the load can be reduced in order to meet power flow constraints post-contingency.
\begin{subequations}
\begin{align}
        &  z_{ijc} = 0       && \forall (i, j) \in \mathcal{L}_c\label{eq:scops_line_contingency}\\
        &  z_{ijc} = z_{ij0}  &&  \forall (i, j) \in \mathcal{L} \setminus \mathcal{L}_c \label{eq:scops_line_active}\\
        &  z_{bc} = z_{b0}  &&  \forall b \in \mathcal{B} \label{eq:scops_bus_active}\\
        &  z_{gc} \le z_{g0}  &&  \forall g \in \mathcal{G} \label{eq:scops_gen_active}\\
        &  x_{dc} \le x_{d0} &&  \forall d \in \mathcal{D} \label{eq:scops_no_pickup}
\end{align}
\label{eq:scops_status}
\end{subequations}
The post-contingency load shed is constrained in equation \eqref{eq:scops_load_shed}, which is similar to the pre-contingency load constraint in \eqref{eq:ops_load_threshold}.  The right hand side has the input parameter $\boldsymbol{\beta}$ between 0 and 1 that determines the portion of load demand that can be shed post-contingency, e.g., $\boldsymbol{\beta}=0.05$ allows 5\% additional load shed to occur post-contingency.  This value multiplies the total load demand in the network $\boldsymbol{D}^{Tot}$ to determine the maximum amount of additional load demand that can be shed in the contingency.  On the left hand side, the total amount of additional load shed from the pre-contingency scenario is determined by taking the difference of $x_{d0}$, the portion of pre-contingency load $d$ that is served, and $x_{dc}$, the portion of load $d$ served in contingency scenario $c$.  This difference represents the increase in load shed in the contingency, and is multiplied by the power demand $\boldsymbol{D}_d$ of each load $d$, and summed over each load. 
\begin{equation}
        \sum_{d\in \mathcal{D}} (x_{d0}-x_{dc})\boldsymbol{D}_d \le \boldsymbol{\beta} \boldsymbol{D}^{Tot}  \label{eq:scops_load_shed}
\end{equation}
The output power of a generator post-contingency $P_{gc}^G$ is limited by the output power in the pre-contingency scenario $P_{g0}^G$ and the generator flexibility input parameter $\boldsymbol{P}_g^{flex} \in [0,1]$, which models the maximum relative percent change in output power shortly after a contingency.  
In Eq. \eqref{eq:scops_gen_ramp}, the output power $P_{gc}^G$ can be at minimum the pre-contingency output $P_{g0}^G$ minus the flexibility multiplied by the generator capacity  $\overline{\boldsymbol{P}_{g}} \boldsymbol{P}_g^{flex}$.  
Similarly, the maximum power output is the pre-contingency scenario power $P_{g0}^G$ plus the flexibility multiplied by the generator capacity  $\overline{\boldsymbol{P}_{g}} \boldsymbol{P}_g^{flex})$.
If the generator is de-energized $z_{gc}=0$  then the output is constrained to $0$. 
\begin{equation}
  z_{gc} (P_{g0}^G \!-\! \overline{\boldsymbol{P}_{g}} \boldsymbol{P}_g^{flex}) \le P_{gc}^G \le z_{gc} (P_{g0}^G \!+\! \overline{\boldsymbol{P}_{g}} \boldsymbol{P}_g^{flex})
   \;\; \forall g \in \mathcal{G} \label{eq:scops_gen_ramp}
\end{equation}

\noindent
\emph{Pre- and Post-Contingency Constraints:} ($\forall c \in \mathcal{C}_0 \cup \mathcal{C}$)
The remaining constraints apply to all pre and post contingency scenarios $c \in \mathcal{C}_0 \cup \mathcal{C}$.  This includes branch power flow, generator power limits, and nodal power balance constraints.
Constraint \eqref{eq:ops_gen_limits} models the generator power. The real power output of generator $P_{gc}^G$ must be between the upper $\overline{\boldsymbol{P}_g}$ and lower $\underline{\boldsymbol{P}_g}$ limits when the generator is energized $z_{gc}=1$ and constrained to $0$ when the generator is de-energized $z_{gc}=0$.
\begin{equation}
        z_{g c}\underline{\boldsymbol{P}_g} \le P_{g c}^G \le z_{g c} \overline{\boldsymbol{P}_g} \quad \forall g \in \mathcal{G} \label{eq:ops_gen_limits}
\end{equation}
The line power flow is modeled in \eqref{eq:ops_flow_limit1}--\eqref{eq:ops_thermal_limit}.
When a line is energized $z_{ijc}=1$ equations \eqref{eq:ops_flow_limit1} and \eqref{eq:ops_flow_limit2} become ordinary DC power flow $P_{i j c}^L = -\boldsymbol{b}_{i j} (\theta_{ic} - \theta_{jc})$.  When a line is de-energized, the big-M value $\boldsymbol{\theta}^{\Delta}_{max}$ allows the power flow $P_{i j c}^L$ to be decoupled from the bus voltage angles $\theta_{i c}, \theta_{j c}$.  This big-M value is calculated as in \cite{hijazi2017convex}. The decoupling allows the power flow to be constrained to $0$ in \eqref{eq:ops_thermal_limit} when a line is de-energized, and within the power limits $\boldsymbol{T}_{i j}$ when energized, without constraining the voltage angles at a bus.
\begin{subequations}
\begin{flalign}
        &  P_{i j c}^L \le -\boldsymbol{b}_{i j} (\theta_{i c} - \theta_{j c} + \boldsymbol{\theta}^{\Delta}_{max}(1-z_{i j c})) && \mkern-20mu\forall (i, j) \in \mathcal{L} \label{eq:ops_flow_limit1} \\
        &  P_{i j c}^L \ge -\boldsymbol{b}_{i j} (\theta_{i c} - \theta_{j c} - \boldsymbol{\theta}^{\Delta}_{max} (1-z_{i j c})) && \mkern-20mu \forall (i, j) \in \mathcal{L} \label{eq:ops_flow_limit2} \\
        & -\boldsymbol{T_{i j}}z_{i j c} \le  P_{i j c}^L \le \boldsymbol{T_{i j}}z_{i j c} && \mkern-40mu \forall (i, j) \in \mathcal{L} \label{eq:ops_thermal_limit}
\end{flalign}
\label{eq:ops_powerflow}
\end{subequations}
The final constraint is \eqref{eq:ops_power_balance} which enforces power balance at each bus.  Here, the total generator power $P_{gc}^G$, line power flow $P_{i j c}^L$, and load $x_{dc}\boldsymbol{D}_d$ must sum to $0$ at each bus in the network. 
\begin{equation}
        \sum_{g\in\mathcal{B}_i^\mathcal{G}}P_{gc}^G - \!\!\!\!\!\! \sum_{(i, j)\in\mathcal{B}_i^\mathcal{L}} \!\!\!\!\!  P_{i j c}^L - \!\!\!\!  \sum_{d\in\mathcal{B}_i^\mathcal{D}} \!\!\! x_{d c} \boldsymbol{D}_d = 0 \; \forall i \in \mathcal{B}  \label{eq:ops_power_balance} 
\end{equation}
\subsection{Security-Constrained Optimal Power Shutoff Model}
The complete optimization model for the SC-OPS problem is given by the following:
\begin{subequations} \label{eq:scops}
\begin{align}
    &\min\limits_{\substack{x, z, P^G, P^L, \theta}} &&  \mbox{${\sum_{(i,j)\in \mathcal{L}} z^L_{i j 0} \boldsymbol{R}^L_{i j}}$ \eqref{eq:ops_objective}} \nonumber\\
&\mbox{s.t.: \,\,\,}  \tag{SC-OPS}
&& \forall c \in \mathcal{C}_0 \nonumber\\
&&& \mbox{Load Delivery:}~\eqref{eq:ops_load_threshold}  \nonumber \\
&&& \mbox{Component relationship:}~\eqref{eq:ops_relationships} \nonumber \\
&&& \forall c \in \mathcal{C} \nonumber \\
&&& \mbox{Contingency Energization:}~\eqref{eq:scops_status} \nonumber \\
&&& \mbox{Contingency Load Shed:}~\eqref{eq:scops_load_shed}  \nonumber \\
&&& \mbox{Generator Response:}~\eqref{eq:scops_gen_ramp} \nonumber \\
&&& \forall c \in \mathcal{C} \cup \mathcal{C}_0 \nonumber \\
&&& \mbox{Generator constraints:}~\eqref{eq:ops_gen_limits} \nonumber\\
&&& \mbox{Power flow constraints:}~\eqref{eq:ops_powerflow} \nonumber \\
&&& \mbox{Power balance constraints:}~\eqref{eq:ops_power_balance} \nonumber
\end{align}
\end{subequations}

\subsection{Optimal Power Shutoff Problem}
In order to study the impact of including security constraints, we also model the OPS problem.  This problem is equivalent to the SC-OPS problem, but with an empty contingency set $\mathcal{C} = \emptyset$. 
\begin{subequations} \label{eq:ops}
\begin{align}
    &\min\limits_{\substack{x, z, P^G, P^L, \theta}} &&  \mbox{${\sum_{(i,j)\in \mathcal{L}} z^L_{i j 0} \boldsymbol{R}^L_{i j}}$ \eqref{eq:ops_objective}} \nonumber\\
&\mbox{s.t.: \,\,\,}  \tag{OPS}
&& \forall c \in \mathcal{C}_0 \nonumber\\
&&& \mbox{Load Delivery:}~\eqref{eq:ops_load_threshold}  \nonumber \\
&&& \mbox{Component relationship:}~\eqref{eq:ops_relationships} \nonumber \\
&&& \mbox{Generator constraints:}~\eqref{eq:ops_gen_limits} \nonumber\\
&&& \mbox{Power flow constraints:}~\eqref{eq:ops_powerflow} \nonumber \\
&&& \mbox{Power balance constraints:}~\eqref{eq:ops_power_balance} \nonumber
\end{align}
\end{subequations}
This model is analogous to the model in \cite{rhodes2020balancing}, but alters the multi-objective by moving load served to a constraint.  This change allows for a direct comparison of the risk of two solutions that serve the same amount of load.

\noindent
{\em OPS Contingency Evaluator:} $(\forall c \in \mathcal{C})$
The OPS problem does not model the system response in contingencies, and modeling this response is necessary to compare the post-contingency load shed of the OPS problem solution to the SC-OPS problem solution. Therefore, we introduce the \emph{Contingency Evaluator} (CE) model to study the post-contingency operation of a scenario. This model evaluates the minimum amount of load shed in a contingency scenario, given a dispatch schedule for the pre-contingency operation with any network configuration. 

In this model each pre-contingency variable is fixed, and treated as a parameter, i.e. $\boldsymbol{z}_{ij0}$, $\boldsymbol{P}_{g0}$ and all other variables where the scenario index $c$ is $0$. 
The objective function of the (CE) problem, shown in \eqref{eq:cv_objective}, minimizes the total load shed $\gamma_c$ that occurs in the post-contingency scenario.
\begin{equation}
    \mbox{\bf minimize: }  \gamma_c\label{eq:cv_objective}
\end{equation}
Constraint \eqref{eq:cv_load_shed} describes the relationship between load shed  in contingencies and variable $\gamma_c$, and is similar to the SC-OPS constraint on post-contingency load shed in \eqref{eq:scops_load_shed}.  The left side of the constraint finds the additional load shed that occurs in contingency $c$ compared to the pre-contingency case by subtracting the percent load served for each load $x_{dc}$ from the pre-contingency case $\boldsymbol{x}_{d0}$ and multiplying this by the power demand $\boldsymbol{D}_d$ at load $d$.   This value is then the amount of additional load shed in the contingency. The right right side multiplies the $\gamma_c$ by the total system load demand $\boldsymbol{D}_d$ to find the limit of post-contingency load shed.  Because  $\gamma_c$ is minimized, this constraint is tight.
\begin{equation}
    \sum_{d\in \mathcal{D}} (\boldsymbol{x}_{d0} \!-\! x_{dc})\boldsymbol{D}_d \le \gamma_c\boldsymbol{D}^{Tot} \;\;  \forall c \in \mathcal{C} \label{eq:cv_load_shed}\\
\end{equation}
Equations \eqref{eq:scops_status}, \eqref{eq:scops_gen_ramp}, \eqref{eq:ops_gen_limits}, \eqref{eq:ops_powerflow}, and \eqref{eq:ops_power_balance} then model the remaining post-contingency constraints of the network.

\section{Case Study} \label{sec:casestudy}

\begin{figure}[t]
    \centering
    \begin{subfigure}{0.7\columnwidth}
        \centering
        \includegraphics[width=1.0\columnwidth]{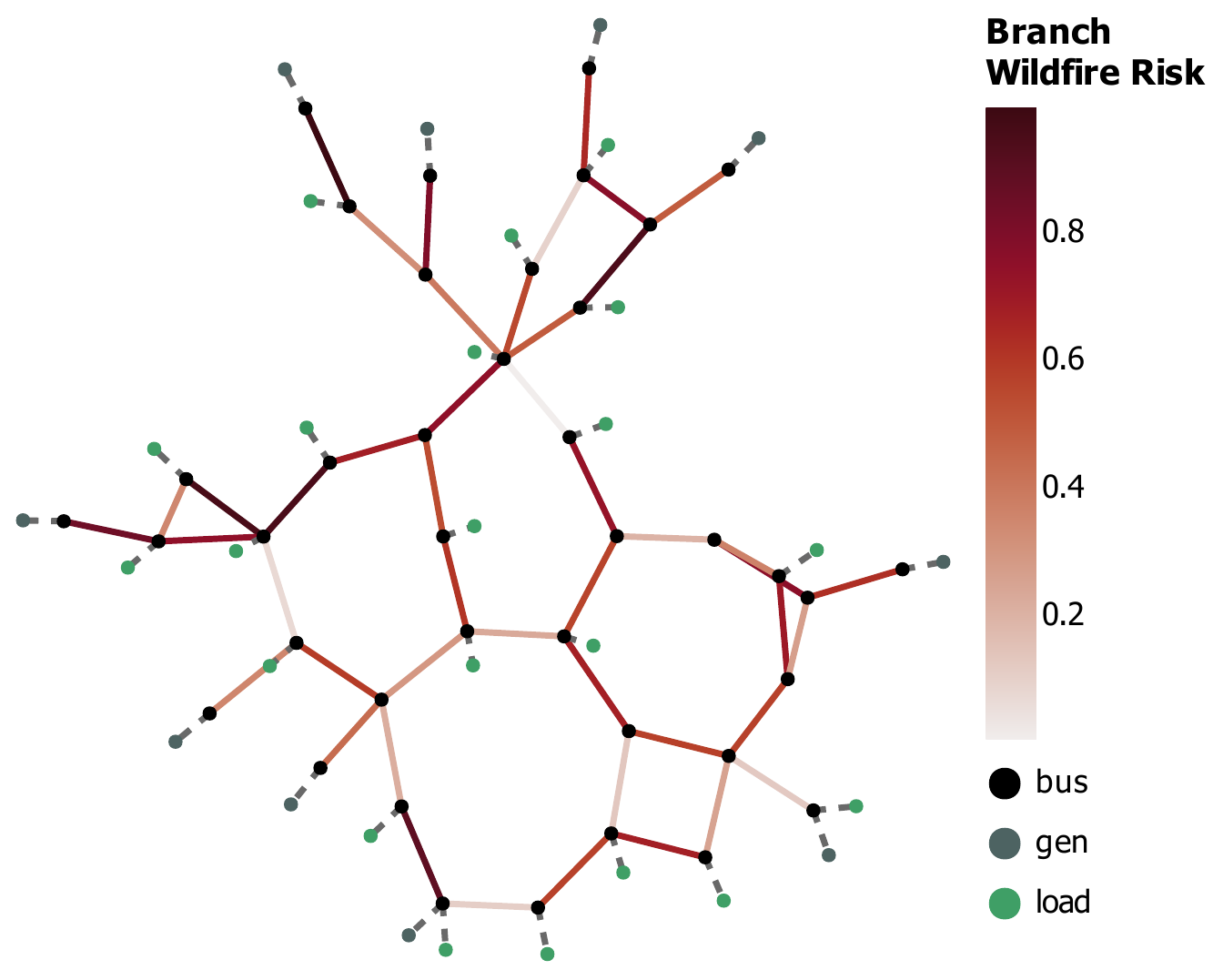} 
        \caption{\small \textbf{Wildfire Risk} } \label{fig:branch_risk}
    \end{subfigure}
    \begin{subfigure}{0.7\columnwidth}
         \centering
        \includegraphics[width=1.0\columnwidth]{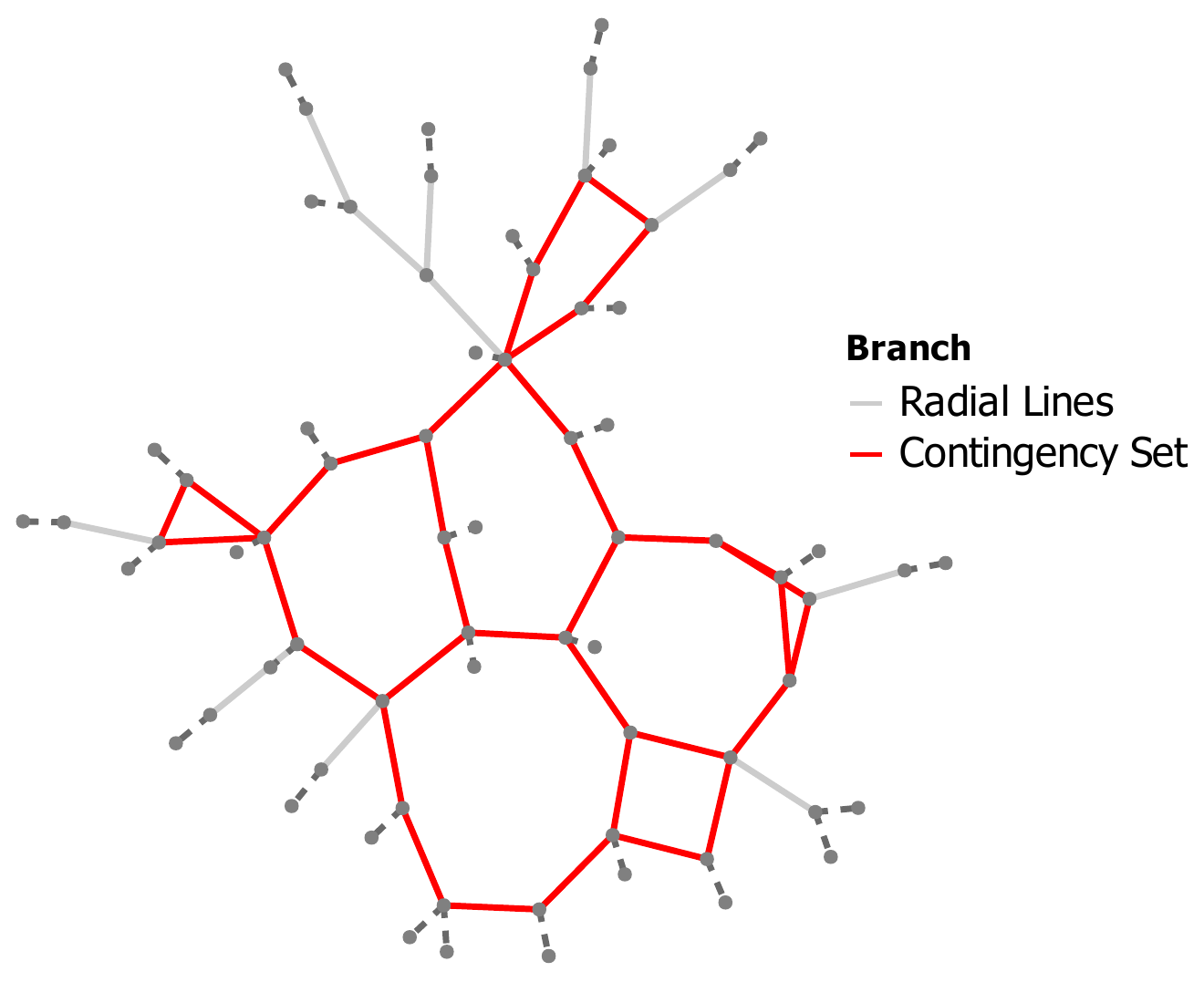} 
        \caption{\small \textbf{Contingency Set} } \label{fig:contingencies}
    \end{subfigure}
    \caption{\small Fig.\ref{fig:branch_risk} shows the wildfire risk of each line. Locations of generators and loads are indicated as grey and green circles and dashed lines indicate where generators and loads connect to the network. Fig. \ref{fig:contingencies} shows each line in the contingency set $\mathcal{C}$ in red. Each grey line is a radial line in the network, and therefore is not considered in the contingency set.}
    \label{fig:casestudy}
    \vspace{-1em}
\end{figure}

In this work, we solve the \emph{Optimal Power Shutoff} and \emph{Security Constrained Optimal Power Shutoff} problems on the Case 39 EPRI test system from PGLib \cite{pglib}.  This 39 bus system has 46 lines, 10 generators, and 21 load buses.  The contingency set $\mathcal{C}$ has 35 N-1 line outage contingency scenarios. Some of the lines in this network are radial, and as a result are not included in the contingency set because they would lead to infeasible solutions.  The network and the set of considered line outages are shown in Fig. \ref{fig:contingencies}.  To demonstrate the model, wildfire risk values for each line are sampled from a uniform distribution from 0 to 1, and shown in Fig. \ref{fig:branch_risk}. In practice, the risk values would be determined from a wildfire risk model such as the Wildland Fire Potential Index (WPFI) \cite{wfpi}, which provides a unit-less value from a pre-defined range to determine the risk. 

The presented models have been implemented in the publicly available software package  \emph{PowerModelsWildfire.jl} \cite{rhodes2020balancing}, and are available for use by researchers and other interested parties. \emph{PowerModelsWildfire.jl} is built using the JuMP mathematical programming language \cite{jump}.  All problems are solved using Gurobi v9.1 \cite{gurobi} with default settings. Power grid figures are generated using \emph{PowerPlots.jl}\cite{powerplots}

\section{Results} \label{sec:results}
We first study the impact of including security constraints by comparing the solutions of the OPS and SC-OPS problems.  Next, we examine how increased system flexibility of loads and generators allows increased risk reduction by decreasing the need for network redundancy. Then we study the impact of allowing increased load shed in the pre-contingency scenarios and how this differs from the impact of load flexibility.  Finally, we present results on the three-way trade-off between pre-contingency load shed, post-contingency load shed, and wildfire risk reduction to demonstrate the range of PSPS options available to operators. 

\subsection{Impact of Adding Security Constraints}

\begin{figure*}[t]
    \centering
    \includegraphics[width=0.85\textwidth]{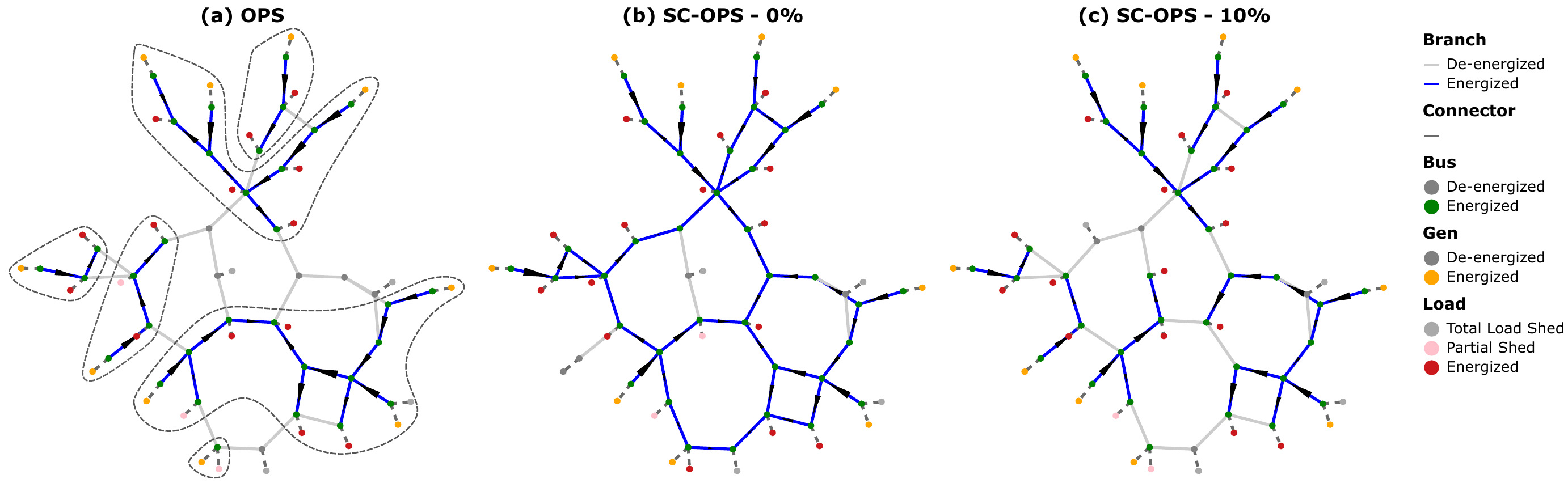} 
    \caption{\small System topology for the OPS solution (right), SC-OPS solution with 0\% additional load shed (middle) and SC-OPS solution with 10\% additional load shed (right). All solutions serve 95\% of the system load demand pre-contingency. 
    The solution to the OPS problem in Fig. \ref{fig:comparison_ops_scops}a results in 6 radial islands in the network (including a single node island), and 51.82\% of the original wildfire risk, but can have up to 17.23\% additional load shed post-contingency.  The solution to the SC-OPS problem with no post-contingency load shed  in Fig. \ref{fig:comparison_ops_scops}b  has many redundant branches energized that increase the risk to 89.23\% of the original wildfire risk. The solution to the  SC-OPS problem with 10\% post-contingency load shed  in Fig. \ref{fig:comparison_ops_scops}c has a similar network structure to the OPS solution and a similar wildfire risk at 53.06\%, but 7.23\% less post-contingency load shed than the OPS problem by co-optimizing pre- and post-contingency load shedding.}
     \label{fig:comparison_ops_scops}
\end{figure*} 

\begin{table}[t]
    \centering
    \caption{\small OPS and SC-OPS solution comparison}
    \resizebox{\columnwidth}{!}
    {\input{tables/solutions_ops_scops_table}}
    \label{tab:comparison_ops_scops1}
    \vspace{-1em}
\end{table}

To demonstrate the benefit of adding security constraints, we compare the solutions of the SC-OPS and OPS problems. For both OPS and SC-OPS, we require $\boldsymbol{\alpha}=95\%$ pre-contingency load served. For the SC-OPS, we allow generator flexibility $\boldsymbol{P}^{flex} = 5\% $ and consider two levels of allowable post-contingency load shed with $\boldsymbol{\beta} = 0\% $ and $\boldsymbol{\beta} = 10\% $, respectively. To evaluate the amount of post-contingency load shed from the OPS solution, we solve the Contingency Evaluator problem with generator flexibility of $\boldsymbol{P}^{flex} = 5\%$.

The solutions for these networks are shown in Fig. \ref{fig:comparison_ops_scops} and Table \ref{tab:comparison_ops_scops1}. 
The OPS solution shown in Fig. \ref{fig:comparison_ops_scops}a contains six islands, all of which are radial, including a single node island.  This is because any redundant power lines add to the risk of wildfire ignition, without supporting additional load delivery.   The wildfire risk is reduced to 51.82\% of the original risk with all branches active while serving the required system load.  When evaluating this shutoff plan with the Contingency Evaluator, we find the worst case post-contingency scenario results in an additional 17.23\% of system load shed.  

Fig. \ref{fig:comparison_ops_scops}b shows the resulting network when solving with the SC-OPS and allowing no post-contingency load shed.  Only five lines are de-energized in this solution, leaving a single island with many redundant power lines to support load in the event of a line outage contingency. However, this added reliability comes at a cost as 89.23\% of the wildfire risk is present on the active lines. 

Permitting 10\% post-contingency load shed results in the SC-OPS solution in Fig. \ref{fig:comparison_ops_scops}c.  This network has 7 radial islands, and has a wildfire risk of 53.06\%, similar to the OPS solution in Fig. \ref{fig:comparison_ops_scops}a.  However, by limiting the post- contingency load shed to 10\%, it has 7.23\% less post-contingency load shed in the worst case contingency compared with the OPS solution.   

Based on these results, we conclude that by using the SC-OPS problem, we can find a shutoff plan with comparable pre-contingency load shed and wildfire risk reduction as the OPS, but with greatly improved post-contingency performance.

\subsection{Impact of Post-Contingency System Response}
\begin{figure*}[t]
    \centering
    \includegraphics[width=1.0\textwidth]{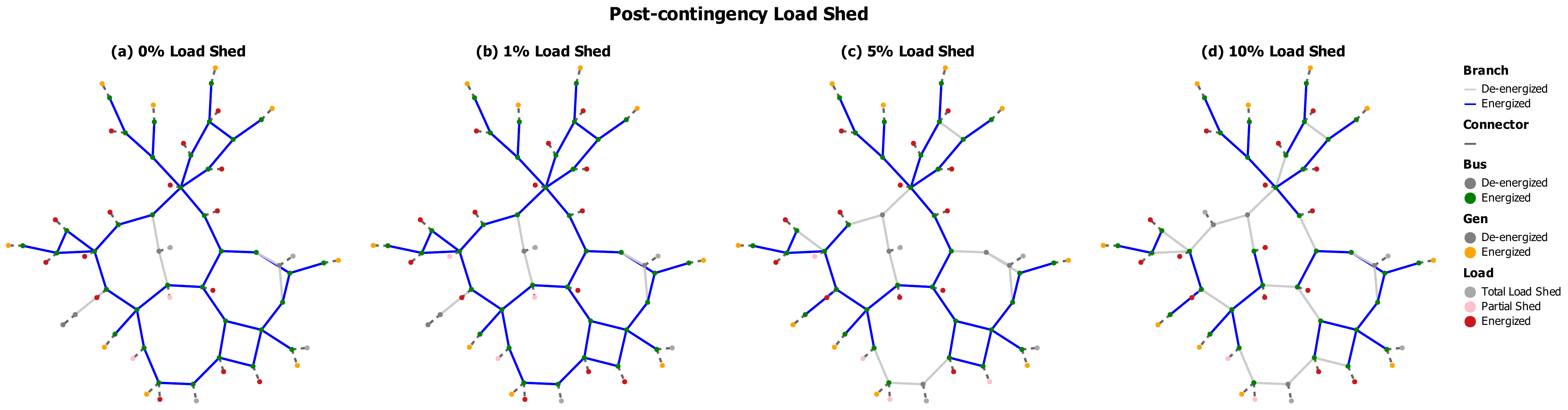} 
    \caption{\small System topology for the SC-OPS solutions with 0\%, 1\%, 5\% and 10\% additional post-contingency load shed (from left to right). Each solution serves 95\% of load pre-contingency, and has 5\% generator flexibility. Moving left to right, we observe that more lines are de-energized and the wildfire risk is lowered as increased post-contingency load shed is permitted. Fig. \ref{fig:shed_system_plot}a has 89.23\% of wildfire risk with 0\% post-contingency load shed,  Fig. \ref{fig:shed_system_plot}b has 89.22\% of wildfire risk with 1\% post-contingency load shed,  Fig. \ref{fig:shed_system_plot}c has 67.51\% of wildfire risk with 5\% post-contingency load shed, and  Fig. \ref{fig:shed_system_plot}d has 53.06\% of wildfire risk with 10\% post-contingency load shed.} 
    \label{fig:shed_system_plot}
    \vspace{-1em}
\end{figure*} 
\begin{figure*}[t]
    \centering
    \includegraphics[width=1.0\textwidth]{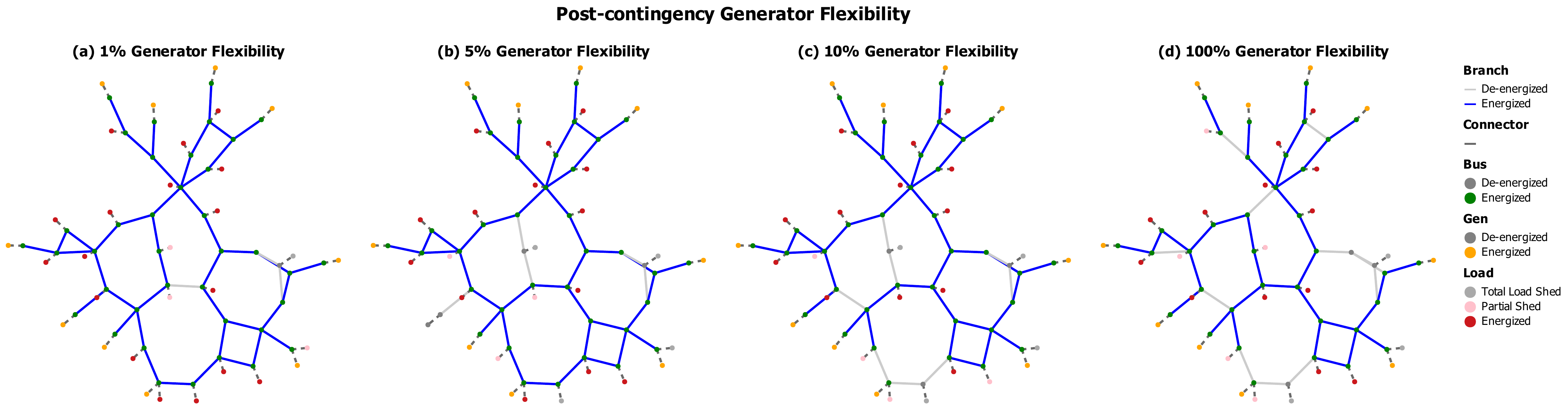} 
    \caption{\small System topology for the SC-OPS solutions with 1\%, 5\% 10\% and 100\% generator flexibility (from left to right). Each solution serves 95\% of system load pre-contingency, and assumes 1\% additional load shed post-contingency. Moving left to right, we observe that more branches are de-energized and wildfire risk is reduced as the post-contingency generator flexibility is increased. Fig. \ref{fig:ramp_system_plot}a has 94.58\% of wildfire risk with 1\% generator flexibility,  Fig. \ref{fig:ramp_system_plot}b has 89.22\% of wildfire risk with 5\% generator flexibility,  Fig. \ref{fig:ramp_system_plot}c has 81.91\% of wildfire risk with 10\% generator flexibility,  Fig. \ref{fig:ramp_system_plot}d has 72.19\% of wildfire risk with 100\% generator flexibility.}
    \label{fig:ramp_system_plot}
    \vspace{-1em}
\end{figure*} 
\begin{figure*}[t]
    \centering
    \includegraphics[width=\textwidth]{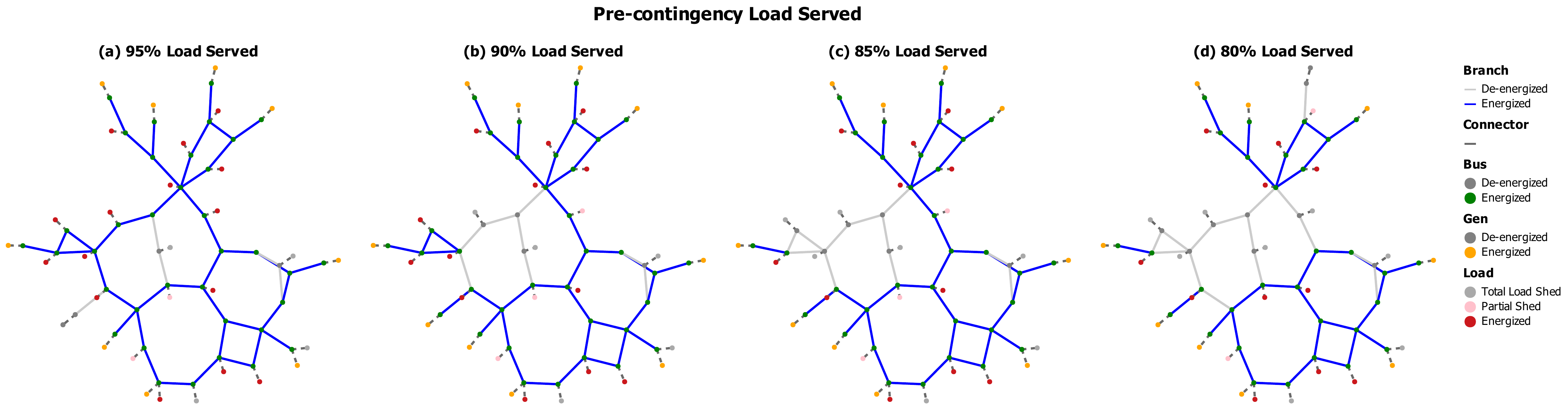} 
    \caption{\small System topology for the SC-OPS solutions with 95\%, 90\% 85\% and 80\% requirement on pre-contingency load served (from left to right). Each solution assumes 0\% additional load shed post-contingency and 5\% generator flexibility. Moving left to right, we observe that reducing the required pre-contingency load results in a large shutoff, mostly contained to a single region. 
    Fig. \ref{fig:threshold_plot}a has 89.23\% of wildfire risk with 95\% pre-contingency load served,  Fig. \ref{fig:threshold_plot}b has 80.64\% of wildfire risk with 90\% pre-contingency load served,  Fig. \ref{fig:threshold_plot}c has 72.16\% of wildfire risk with 85\% pre-contingency load served,  Fig. \ref{fig:threshold_plot}d has 63.99\% of wildfire risk with 80\% pre-contingency load served. }
     \label{fig:threshold_plot}
     \vspace{-0.5em}
\end{figure*} 
Adding security constraints requires the power network to have more redundancy to ensure continued operation after a contingency. This redundancy can be achieved by having redundant transmission paths (which leads to a meshed network topology, as observed above). However, in the context of power shut-offs, having redundant paths means that more lines must remain energized, thus increasing the wildfire risk. An alternative to redundant transmission paths is to allow greater flexibility of generation and load. More flexible generation capacity allows local generators to increase/decrease output power in case of a contingency, while load flexibility can be achieved by accepting a larger amount of post-contingency load shed. We next investigate how added load and generation flexibility impact the achievable reduction in wildfire risk for a given level of pre-contingency load shed.

\subsubsection{Impact of allowing post-contingency load shed}
We first analyze the topology change observed in the SC-OPS solutions as the permitted post-contingency load shed is increased from 0\% to 10\% while 95\% of load is served in the pre-contingency scenario and the generation flexibility is kept constant at 5\%. The results are shown in Fig. \ref{fig:shed_system_plot}a-d. We see that allowing increased post-contingency load shed permits more branches to be de-energized because the network is no longer required to support the full load and associated power flows.

\subsubsection{Impact of post-contingency generation flexibility}
Next, we analyze how the topology changes as the generator flexibility is changed from 1\% to 100\%. We use a post-contingency load shed limit of 1\% and require 95\% of load to be served in the pre-contingency scenario. 
The results are shown in Fig. \ref{fig:ramp_system_plot}a-d. We see that as the flexibility increases, more components are de-energized, the network becomes more radial, and the wildfire risk is reduced.
If generator flexibility can reach 100\% of maximum output, i.e. a 100\% inverter based system with solar and batteries, then the generator flexibility allows almost all redundant lines to be de-energized while ensuring that the post-contingency load shed never exceeds 1\% of system load. 

From these results, we observe that either allowing more load shed in contingency scenarios or increasing the generator flexibility both amount to relaxing the SC-OPS problem, and both encourage a sparser and more radial system topology. This allows higher wildfire  risk reduction without reducing the amount of pre-contingency load served. 
However, a challenge with permitting additional load shed in contingencies is that customers may have uncertain access to electricity.  Being notified of a planned shutoff allows them to plan for a sustained outage, whereas post-contingency outages make the uncertainty 
a challenge for customers.

\subsection{Impact of Allowing More Pre-Contingency Load Shed}

The last main parameter in designing a PSPS is the amont of allowed pre-contingency load shed. This parameter is different from the post-contingency system response discussed above, because customers are directly impacted, regardless of whether an additional contingency takes place. %
To assess how the allowable amount of pre-contingency load shed impacts the solution, we solve the SC-OPS problem with a generator flexibility of 5\% and post-contingency additional load shed of 0\%.  The required pre-contingency load served is 95\%, 90\%, 85\% and 80\% in each scenario. 

Fig. \ref{fig:threshold_plot} shows the network structure for these four scenarios. The de-energized lines all expand from the first two de-energized lines in center-left section of the 95\% load served scenario, creating a large de-energized region as the load shed is increased.
We note that increasing the amount of allowed pre-contingency load shed has a very different impact on the network structure of the solution compared to allowing higher post-contingency load shed or increasing generator flexibility.  While increasing post-contingency load shed or generator flexibility promoted a sparser and less redundant network to reduce wildfire risk,
increased pre-contingency load shed maintains a highly redundant meshed network but with several fully de-energized regions.
Even in the scenario which requires only 80\% of pre-contingency load to be served there are still four cycles in the network, providing redundant routes for power flow in the event of a contingency.  
This is required since the post-contingency response of generators and loads is limited, and continued serving of load must be accommodated through network redundancy.

We note that allowing increased pre-contingency load shed still results in higher wildfire risk than allowing higher post-contingency load shed.  In the scenario that allows $20\%$ pre-contingency load shed (i.e. requires only 80\% of the pre-contingency load to be served) and $0\%$ load shed post contingency, the wildfire risk is $63.99\%$ of the original risk. In comparison, the scenario which allows $5\%$ load shed in the pre-contingency scenario (i.e. requires 95\% of pre-contingency load to be served) and $10\%$ additional post-contingency load shed, reduces the wildfire risk to $53.06\%$ of the original risk.  

\subsection{Impact of Allowable Pre- and Post-Contingency Load Shed on Wildfire Risk Reduction}

We further investigate the pre- and post-contingency load shed limits impact the achievable wildfire risk reduction in the SC-OPS, and compare these results with the OPS solution. 
\subsubsection{Wildfire Risk Reduction for Different Pre- and Post-Contingency Load Shed Limits} Selecting a 'correct' trade-off of wildfire risk and and load shed is impossible without the ability to quantify the cost of wildfire risk, but our results show the relative impact of the two parameters, and how they can combine to reduce risk.

\begin{figure}[t]
    \centering
        \includegraphics[width=0.9\columnwidth]{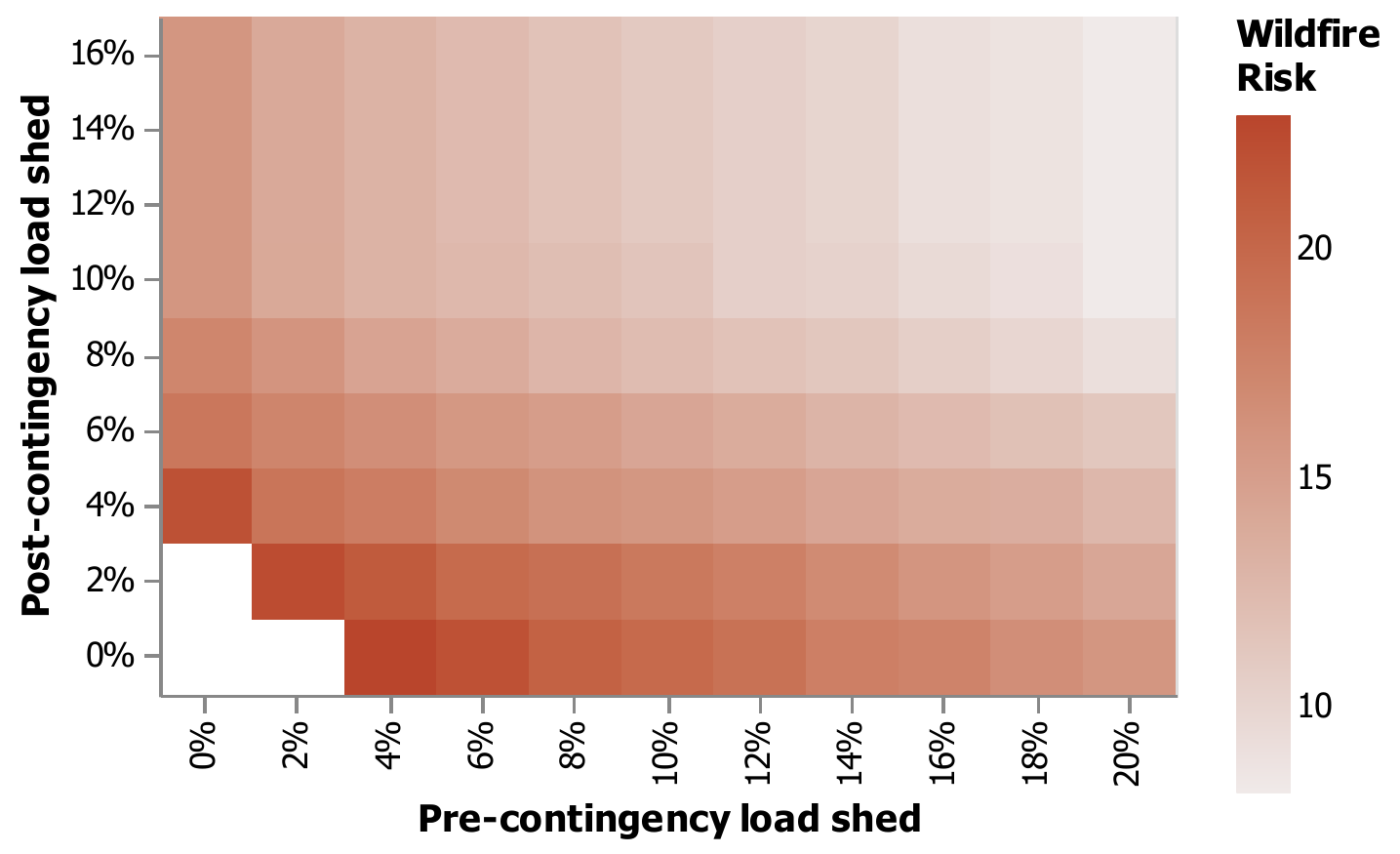}
    \caption{\small \textbf{Wildfire Risk for Varying Levels of Pre- and Post-Contingency Load Shed:} This figure shows the wildfire risk of a solution for different pre-contingency load shed limits (horizontal axis) and post-contingency load shed (vertical axis). The wildfire risk is indicated by the color of the corresponding pixel in the figure, with darker red indicating higher risk. 
    }
    \label{fig:risk_heatmap}
    \vspace{-1em}
\end{figure}

\begin{figure}[t]
    \centering
    \begin{subfigure}{0.9\columnwidth}
        \centering
        \includegraphics[width=\columnwidth]{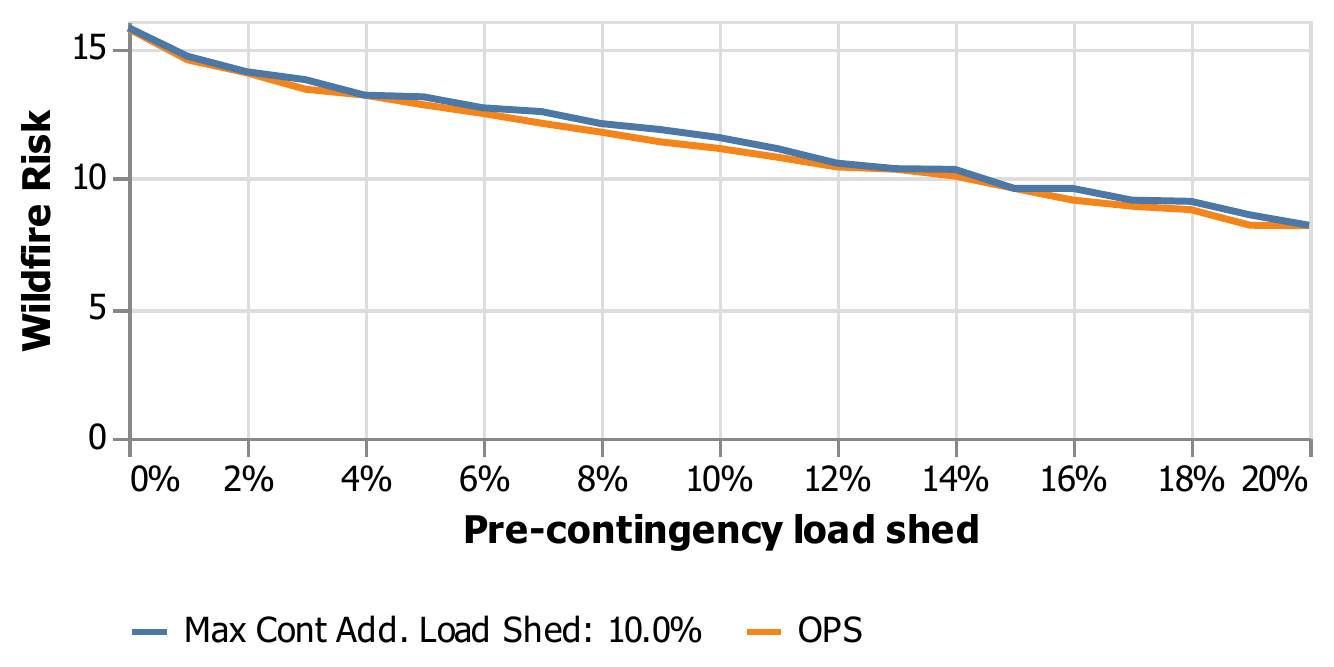}
        \caption{\small \textbf{Wildfire Risk vs. Load Served }}
        \label{fig:tradeoff_loadramp}
    \end{subfigure}
    \begin{subfigure}{0.9\columnwidth}
        \centering
        \includegraphics[width=\columnwidth]{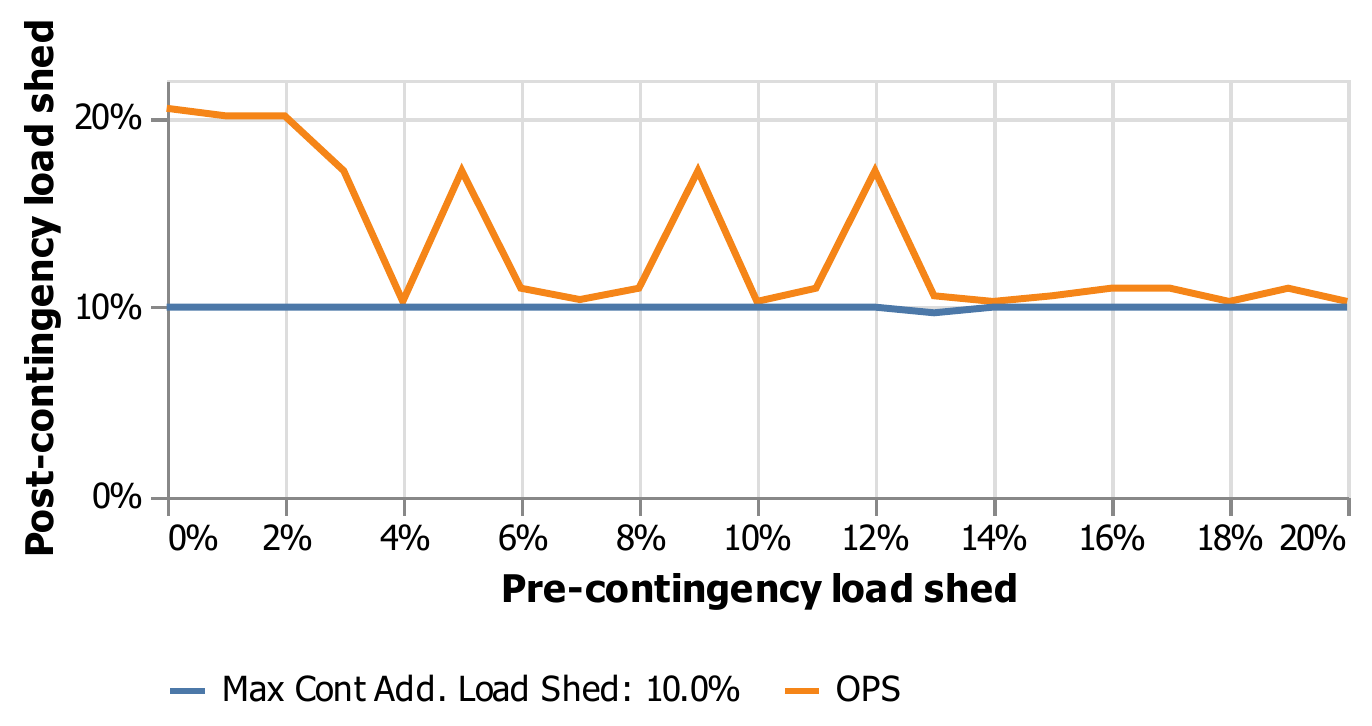}
        \caption{\small \textbf{Pre-Contingency Load Shed vs.  Post-Contingency Load Served}}
        \label{fig:tradeoff_loadcont}
    \end{subfigure}
    \caption{\small Fig. \ref{fig:tradeoff_loadramp} shows the wildfire risk for a given value of load served for the OPS problem (in orange) and for the SC-OPS problem with 10\% contingency load shed (in blue).  %
    Fig. \ref{fig:tradeoff_loadcont} shows the amount of post-contingency load shed  for a given value of load served for the OPS problem (in orange) and for the SC-OPS problem with 10\% contingency load shed (in blue).
    }
    \label{fig:tradeoffs}
    \vspace{-1em}
\end{figure}

We solve the SC-OPS problem with a generator flexibility of 5\%, pre-contingency load shed ranging from 0\% to 20\% in steps of 2\% (corresponding to requiring that between 100\% and 80\% of all pre-contingency load is served), and post-contingency load shed ranging from 0\% to 16\% in steps of 2\%.  The wildfire risk for each solution is shown in Fig. \ref{fig:risk_heatmap}.
Each pixel in the plot represents a solution to the SC-OPS problem for a given limit on pre- and post-contingency load shed, and is shaded according to the wildfire risk of the solution.
Three solutions are infeasible, and shown in white in the bottom left.  This figure shows that increasing either the permitted amount of pre-contingency load shed and post-contingency load shed allows for solutions with lower wildfire risk, 
and allowing a combination of both pre- and post-contingency load shed is necessary to obtain the lowest risk solution. 

It is interesting to observe that for a given maximum percentage load shed (e.g. 10\%), the resulting wildfire risk is lower if the load shed is performed post-contingency. This is consistent with our finding from the previous section. It can be explained by the fact that post-contingency load shed is targeted specifically to the contingency that took place (and can differ across contingency scenarios) whereas pre-contingency load shed has to be shared across all possible contingencies. Thus, post-contingency load shed allows for a more flexible response. 
We conclude that allowing post-contingency load shed is beneficial both for reducing wildfire risk and to reduce the probability that customers may experience outages (as it is relatively unlikely that a specific contingency will take place). The main drawback of post-contingency load shed is that it happens suddenly and without warning. 

\subsubsection{Comparison with OPS}
We next compare this range of solutions to the solution obtained with the OPS problem.
We first solve the SC-OPS problem with 5\% generator flexibility and 10\% post-contingency load shed, while the pre-contingency load shed limit is varied from 0\% to 20\% in steps of 1\%.  The OPS problem is solved with the same limits on pre-contingency load shed. The post-contingency load shed for the OPS problem  is evaluated using the \emph{Contingency Evaluator}, which minimizes the amount of post-contingency load shed given a fixed PSPS plan using a generator flexibility of 5\%. For these SC-OPS and OPS solutions, we compare the resulting wildfire risk and post-contingency load shed.

Fig. \ref{fig:tradeoff_loadramp} shows the wildfire risk as the pre-contingency load shed is increased from 0\% to 20\%. 
We observe that the SC-OPS solution with 10\% additional post-contingency load shed results in solutions with similar wildfire risk as the OPS problem across the range of pre-contingency load shed limits.  

Fig. \ref{fig:tradeoff_loadcont} shows the post-contingency load shed of the OPS and SC-OPS solutions for pre-contingency load shed limits ranging from 0\% to 20\%. The SC-OPS solutions all have post-contingency load shed close to 10\%, which indicates the constraint on post-contingency load shed is often tight. In contrast, the post-contingency load shed of the OPS solutions varies from 10.3\% to 20.5\%. 

These results further demonstrate the value of incorporating security constraints in the power shut-off planning. 
The OPS solution represents a lower bound on the achievable wildfire risk reduction, as it excludes the contingency constraints. At the same time, we observe that the post-contingency load shed is consistently lower in the SC-OPS solution compared with the OPS solution. We can thus conclude that for the case study discussed here, solving the SC-OPS with 10\% additional post-contingency load shed allows us to obtain a PSPS plan with comparable wildfire risk, but lower vulnerability to contingencies compared with solving only the OPS problem. 

\subsection{Solve Time}
While solving the optimization problems to create Figs. \ref{fig:risk_heatmap} and \ref{fig:tradeoffs}, we noted the wide range of solution times for the OPS and SC-OPS problems.  The OPS problem solution time ranged from 0.2 seconds to 2.95 seconds as the input parameter for pre-contingency load shed was altered.  The SC-OPS solution time ranges from 11.35 seconds to 1774 seconds as both the pre-contingency and post-contingency load shed limits are changed. This means that the input parameters can have an over 100x impact on the solution time of this type of shutoff problem.  Adding security constraints to this small 39-bus network also has a wide impact on solution time as the SC-OPS problem ranges from 6.9x to 8,673x slower than the OPS problem for the same pre-contingency load shed parameter.

This SC-OPS formulation was not designed to be the fastest method to solve the SC-OPS problem.  More sophisticated mathematical programming methods could be used to improve the solution time, and in particular reduce the variance in solution time as the input parameters for load shed are modified.  This is an avenue for future work that is required for this optimization problem to scale to realistic sized networks.

\section{Discussion} \label{sec:discussion}
The results presented above demonstrate several important aspects of PSPS planning that we want to highlight.

First, our results clearly demonstrate the benefit of considering the impact of contingencies when planning a PSPS.
Incorporating security constraints in the SC-OPS formulation allows utilities to limit the size of additional post-contingency power outages, thus achieving solutions with lower post-contingency outages at similar pre-contingency load shed and wildfire risk compared with the OPS solutions where such impacts are not accounted for.

Second, compared to traditional security-constrained OPF which requires post-contingency load shed to be zero, we observe that the SC-OPS exhibits significant benefits of allowing load shed post-contingency. Specifically, if we allow a given percentage of load to be shed across the pre- and post-contingency operating conditions, we achieve the largest reduction in wildfire risk if most of this load shed happens post-contingency. This is because the load can be shed based on the specific contingency.

The fact that post-contingency load shed is more effective than pre-contingency load shed has a few important implications. First, post-contingency load shed only happens if a contingency takes place, which is a low-probability event. This is in stark contrast to pre-contingency power shut-offs, which more or less guarantee power outages to customers. Furthermore, the SC-OPS problem only limits the worst-case post-contingency load shed, and the load shed will be lower in most contingency situations. Considering both the low probability that a contingency happens and the fact that most contingencies will have smaller load shed than the allowable limit, the \emph{expected} (i.e. probability-weighted) power outages experienced by customers is much smaller when utilities rely on post-contingency rather than pre-contingency load shed.

Third, if utilities choose to rely on post-contingency load shed, they are still exposing their customers to a higher-than-normal risk of power outages as a single contingency is enough to cause load shed (as compared to standard N-1 secure operation). It is therefore important to consider how to best communicate with and possibly notify customers regarding this heightened risk. Utilities in wildfire-prone areas have systems in place to notify customers about upcoming shut-offs, and similar systems could likely be used to provide warnings about a higher risk of outages. Notification of post-contingency load shedding is not addressed in this work, but is an important
feature for implementation of PSPS.

Fourth, our results demonstrate the benefits of flexible generation to manage changes in the power flow post-contingency and continue to serve customers.

\section{Conclusion} \label{sec:conclusion}
Public safety power shutoffs (PSPS) are a method to reduce the risk of wildfire ignitions by de-energizing components in the power grid. This wildfire risk reduction comes at the cost of power outages to customers. Furthermore, as discussed in this paper, PSPS impact reliability of the network and increase the probability that additional unplanned contingencies such as line failures will cause additional load shed. 

To address this problem, we presented the Security-Constrained Optimal Power Shutoff problem. This optimization problem minimizes system-wide wildfire risk and load-shed, and extends previous work by incorporating security constraints (i.e. constraints that model system operation after a contingency) into the problem.  
Specifically, our formulation limits the worst-case post-contingency load shed.

The case study in the work demonstrates the trade offs between load shed, wildfire risk, and system reliability. We observe that it is necessary to allow some post-contingency load shed, otherwise the incorporation of security constraints reduces the amount of wildfire risk reduction that is possible. 
Relaxing the problem (permitting some load shed in contingencies) allows solutions with comparable wildfire risk and pre-contingency load shed as the OPS problem (without security constraints), while at the same time achieving lower post-contingency load shed (i.e., increased system reliability). This demonstrates the benefits of explicitly considering post-contingency operation in PSPS planning. 
Further, our case study shows that increasing generator flexibility allows us to achieve an improved wildfire risk reduction for the same amount of load shed. 

While it is beneficial to consider security constraints, they also add increased computational burden to an already difficult to solve mixed-integer optimization problem. Future work may require different approaches to address reliability requirements of PSPS problems in a computationally tractable way for realistic scale power grids. Other avenues for future work include verification of AC power flow feasibility of the solutions, as well as consideration of customer notification times.

 \balance
\bibliographystyle{IEEEtran}
\bibliography{IEEEabrv,references}

LA-UR-23-24226

\end{document}

%% file: tables/solutions_ops_scops_table.tex
\begin{tabular}{cccc} \\ %
\multicolumn{1}{p{1.8cm}}{\centering Scenario: Problem Model}
 & \multicolumn{1}{p{1.3cm}}{\centering Base Load Served \% }
 & \multicolumn{1}{p{1cm}}{\centering Active Risk \% }
 & \multicolumn{1}{p{1.8cm}}{\centering Max Cont Add. Load Shed \% }
\\ \hline
SCOPS - 0\%  & 95.00\% & 89.23\% & 0.00\% \\ %
SCOPS - 10\%  & 95.00\% & 53.06\% & 10.00\% \\ %
OPS  & 95.00\% & 51.82\% & 17.23\% \\ %
\end{tabular}